\documentclass[pre,twocolumn,a4paper,showpacs]{revtex4-1}
\usepackage{amsmath} 
\usepackage{wasysym}
\usepackage{graphicx} 
\usepackage{textcomp} 
\usepackage{subfigure} 
\usepackage[table]{xcolor} 
\usepackage{amssymb} 
\usepackage{amsfonts} 
\usepackage{appendix}

\begin{document}

% \runauthor{Paulo R. A. Campos}
% \runauthor{Jos\'e F. Fontanari }

\title{Predictability of the imitative learning trajectories}                                                                                                                                  

\author{Paulo R. A. Campos}
\affiliation{Departamento de F\'{\i}sica, Universidade Federal de Pernambuco,
52171-900 Recife, PE, Brazil}

\author{Jos\'e F.  Fontanari}
\affiliation{Instituto de F\'{\i}sica de S\~ao Carlos, Universidade de S\~ao Paulo, Caixa Postal 369, 13560-970 S\~ao Carlos, S\~ao Paulo, Brazil}

\begin{abstract}
The fitness  landscape metaphor plays a central role on the modeling of  optimizing principles in many research fields, ranging from evolutionary biology, where it was first introduced, to management research. Here we consider the ensemble of  trajectories of  the imitative learning search, in which agents exchange information on their fitness and  imitate the fittest  agent in the population aiming at reaching the global maximum of the fitness landscape. We  assess the degree to which the starting and ending points determine the learning trajectories using two measures, namely, the predictability that yields the probability that two randomly chosen   trajectories  are  the same, and  the mean path divergence that  gauges  the dissimilarity between  two  learning trajectories. We find that the predictability is greater in rugged  landscapes than in smooth ones. The mean path divergence, however, is strongly affected by the  search parameters -- population size and imitation propensity -- that obliterate the  influence of the underlying landscape. The  learning trajectories become more deterministic, in the sense that there are fewer distinct trajectories and those trajectories are more similar to each other, with increasing population size and  imitation propensity. In addition, we find that the roughness of the learning trajectories, which measures the deviation from additivity of
the fitness function,  is always greater than the  roughness estimated over the entire fitness landscape.
\end{abstract}

%keywords: optimization, predictability, NK landscapes

\maketitle

\section{Introduction}\label{sec:intro}

Search is fundamental for human  decision-making because  the available choices  usually result from discovery  through search activities \cite{Simon_71}. Hence,  given the  central role played by  imitation on the construction of  human culture   \cite{Blackmore_01, Boyd_10}, which was neatly  summarized  by the  phrase ``Imitative learning acts like a synapse, allowing information to leap the gap from one creature to another'' \cite{Bloom_01},  it has been suggested that agent-based models of the imitative learning search could reproduce some features of  problem-solving  strategy and performance of groups \cite{Kennedy_99,Lazer_07,Fontanari_14,Fontanari_15}. 
In fact, in the case  that the agents are too propense to imitate their more successful peers or that the group is too large,  the  resulting  disastrous performance of the imitative learning search, as  compared with the baseline situation where  the agents work independently of each other,  is reminiscent of   the classic Groupthink phenomenon of social psychology that occurs when everyone in a group starts 
thinking alike \cite{Janis_82}.
Otherwise, if the imitation propensity and the group size are properly set, then  the imitative search greatly improves the group performance, as expected  \cite{Fontanari_15}.

The standard theoretical framework to study  search heuristics is the fitness landscape metaphor  first introduced in the context of evolutionary biology  to assess the consequences of the genotype-fitness map \cite{Wright_32}. The basic idea is that points in a multidimensional space representing  all possible gene combinations (genotypes) compose the domain of a real valued function that represents the fitness of each genotype. Since the expected effect of natural selection is the increase of fitness, evolution can be viewed as a hill-climbing or adaptive walk in the space of genotypes.  The  effectiveness of natural selection would thus be strongly dependent on the ruggedness of the fitness landscape  \cite{Carneiro_10}.

Within the fitness landscape framework we can, in principle, represent the entire evolutionary  history of a population by a trajectory on the landscape, and so  the analysis of the ensemble  of the  evolutionary trajectories can offer an approach to the  (open) issue of whether evolution is predictable or not.   In other words, if  ``the tape of evolution is replayed" should we expect a completely different outcome? We refer the reader to  \cite{Gould_97}  and   \cite{Morris_10} for different viewpoints on this matter. 
%Of course,  the issue here is not whether the population dynamics is deterministic or stochastic, but whether the  constraints imposed by the underlying fitness landscape are  strong enough to drive the (stochastic) evolution towards the same evolutionary paths.

The  study of evolutionary predictability  through the analysis of the ensemble of evolutionary trajectories resulted in the proposal of (at least) two measures  aiming at quantifying the diversity and divergence of  trajectories that  begin at a same point in genotype space and end also at a same, distinct point, usually  the wild type of a gene that codifies an enzyme \cite{Lobkovsky_11,Lobkovsky_12,Visser_14}.  The first measure is the predictability that yields the probability that two evolutionary  trajectories picked at random from the ensemble of trajectories are  the same.  The second measure is the mean path divergence that  gauges  the dissimilarity between  two randomly chosen trajectories in the ensemble of trajectories.
Those measures  indicate that adaptive walks in rugged landscapes are more deterministic and hence more predictable  than walks in smooth   landscapes. However, due mainly to the technical difficulty of defining and following  evolutionary  trajectories using  standard evolutionary algorithms,  there are no systematic studies of the effects of different degrees of ruggedness on the statistical properties of the ensemble of trajectories.

Here we consider the ensemble of learning trajectories of  a well-mixed population of agents that search for the unique global maximum of NK-fitness landscapes \cite{Kauffman_87,Kauffman_95} using imitative learning. In this context, there is little ambiguity in the definition of a learning trajectory, which is the ordered sequence of states assumed by the model agent, i.e., the fittest agent  in the population, who can influence all the other agents through  the imitation procedure. We note that the NK model is the canonical modeling approach used in management research to study decision making in complex scenarios where the decision variables interact in determining the value  (fitness) of their combinations \cite{Levinthal_97,Billinger_13, Sebastian_17}.
In addition, use of the NK model allows the tuning of the ruggedness of the landscape and hence the study of the  effects of the topography of the  fitness landscape on the  repeatability of the imitative learning trajectories in distinct searches.

We find that the properties of the ensemble of learning trajectories are determined by the interaction between the  search dynamics and the
 underlying fitness landscape. In agreement with the results for evolutionary trajectories, we find that the learning trajectories are more deterministic, in the sense that they are more similar to one another, for rugged landscapes   than for  smooth (additive)  ones. However,  increasing the ruggedness of the landscape, say by increasing the parameter $K$ of the NK model, does not necessarily make the search more deterministic. 
 The distance and the difference in fitness between the local maxima are probably  influential factors to their attractivity, which greatly impact  on the properties of the learning trajectories. 
 The increase of either the population size or the imitation  propensity  makes the search more deterministic since both parameters enhance the attractivity of the local maxima, thus forcing the  trajectories to pass through them. The existence of few escape routes from the local maxima results in more predictable and less divergent   learning trajectories.
 
Another interesting aspect of the study of trajectories, which is not directly related to the predictability  issue, is the possibility of assessing the roughness, defined as the deviation from additivity \cite{Aita_01}, experienced by the search on a specific learning trajectory. This analysis shows that changing the parameters of the model, i.e.,  the population size $M$ and the imitation propensity $p$,  changes the  characteristics
of the trajectories  of the imitative search. In particular, the roughness of the learning trajectories increases monotonously with increasing   $M$ and $p$, and tends to a well-defined value for $M \to \infty$ which corresponds to the trajectory of maximum roughness in the landscape. In addition, we find that the learning  trajectories  exhibit a  roughness much greater than the average roughness of the landscape.

The rest of this paper is organized as follows. In Section \ref{sec:NK} we offer  an outline of the NK model of   rugged fitness landscapes  \cite{Kauffman_87} and present a measure of ruggedness,  denoted by  roughness \cite{Aita_01},  that we use to characterize those landscapes. The  rules used by the agents to explore the state space of the NK-landscape  are explained in Section \ref{sec:model} together with the definition of the  (purged) learning trajectories.  The measures of  predictability and similarity of the learning trajectories are then introduced in Section \ref {sec:predic}.
In  Section \ref{sec:res} we present and analyze the results of our simulations, emphasizing the effect of the ruggedness of the landscape on the learning trajectories.  Finally, Section \ref{sec:disc} is reserved for our concluding remarks.

\section{NK model and roughness}\label{sec:NK}

The NK model \cite{Kauffman_87} is the choice computational implementation  of fitness landscapes that 
has been extensively used to study optimization problems in population genetics, developmental biology and protein folding \cite{Kauffman_95}. In fact, the repute of the NK model goes way beyond the (theoretical) biology realm, as that model is considered
a paradigm  for problem representation in management research  \cite{Levinthal_97,Billinger_13,Sebastian_17},
 since it allows the manipulation of the  difficulty of the problems and challenges posed to  individuals and companies.
The NK model of rugged fitness  landscapes is defined in the space of binary strings of length $N$ and so this parameter determines the size of the state space or the dimensionality of the landscape, namely, $2^N$.   The other parameter,  $K =0, \ldots, N-1$, determines the range of the epistatic interactions among the bits of the binary string and  influences strongly the number of local maxima on the landscape.

More pointedly, the state space of the NK landscape consists of the $2^N$ distinct binary strings of length $N$, which we denote by $\mathbf{x} = \left ( x_1, x_2, \ldots,x_N \right )$ with
$x_i = 0,1$. To each string $\mathbf{x}$ we associate a fitness value $\mathcal{F} \left ( \mathbf{x}  \right ) $ that  is given by an average  of the contributions from each  component $i$ of the string, i.e.,
\begin{equation}\label{NK_1}
\mathcal{F} \left ( \mathbf{x}  \right ) = \frac{1}{N} \sum_{i=1}^N f_i \left (  \mathbf{x}  \right ) ,
\end{equation}
where $ f_i$ is the contribution of component $i$ to the  fitness of string $ \mathbf{x}$. The local fitness  $ f_i$ depends on the state $x_i$  as well as on the states of  $K$  distinct randomly chosen components, i.e., $f_i = f_i \left ( x_i, x_{i_1}, \ldots, x_{i_K} \right )$, with $i_1 \neq i_2 \ldots \neq i_K \neq i$, so each $f_i$ has $2^{K+1}$ distinct arguments. It is clear then that the parameter $K$ determines the degree of interaction (epistasis) among the components of the string. As usual,  we assign a uniformly distributed random number  in the unit interval to each one of the $2^{K+1}$ distinct arguments of $f_i$, so the specification of the full NK fitness landscape requires the generation of $N 2^{K+1}$ uniform deviates \cite{Kauffman_87}.  As a result, we have  $\mathcal{F}  \left ( \mathbf{x}  \right )  \in \left ( 0, 1 \right )$ for all strings $\mathbf{x}$. Moreover, because  the  fitness  $\mathcal{F} \left ( \mathbf{x}  \right )$ are real valued random variables with support in the unit interval,   the NK landscape  exhibits a unique global maximum   with probability one, as  different strings  have different fitness values.

For $K=0$, the NK-fitness landscape  exhibits a single maximum, which is easily determined by picking for each component $i$ the state $x_i = 0$ if  $f_i \left ( 0 \right ) >  f_i \left ( 1 \right )$ or the state  $x_i = 1$, otherwise. In time, a string is a maximum if its fitness is greater than the fitness of 
all its $N$ neighboring  strings (i.e., strings that differ from it at a single component).
For $K=N-1$, the fitness values  
of neighboring configurations    are  uncorrelated and so the NK model reduces to the Random Energy model \cite{Derrida_81,Saakian_09}.  This (uncorrelated) landscape  has on the average  $2^N/\left ( N + 1 \right)$ maxima with respect to single bit flips \cite{Kauffman_87}.

A simple and popular measure of the ruggedness of a landscape  is the density of  maxima, i.e., the number of strings with no fitter neighbors divided by the total number of strings in the landscape \cite{Kauffman_87}. However, here we will  consider   an alternative  measure  that has been used to  model empirically adaptive landscapes in the  protein evolution literature \cite{Aita_01,Carneiro_10,Lobkovsky_11,Visser_14}:  the roughness  $\rho$ that measures the  deviation from additivity of a landscape. Since $\rho$ is measured  over a particular trajectory towards the global maximum, it is a valuable tool to assess the nature of the search on  rugged landscapes. In the following we offer a brief sketch of this important measure \cite{Aita_01}.

Without loss of generality, we can assume that the fitness $\mathcal{F} \left ( \mathbf{x}  \right )$ of a string $\mathbf{x}$ is given by the  sum of an additive term $F \left ( \mathbf{x}  \right )$ and a non-additive term $\Omega \left ( \mathbf{x}  \right )$, i.e.,
\begin{equation}\label{NK_2}
\mathcal{F} \left ( \mathbf{x}  \right ) = F \left ( \mathbf{x}  \right ) + \Omega \left ( \mathbf{x}  \right ) .
\end{equation}
Ideally, the non-additive term  $\Omega$  should be a small random component because, in practice, protein fitness landscapes are nearly additive close to the maximum (wild type)  \cite{Carneiro_10},  but here we impose no condition  on $\Omega$ since   $\mathcal{F} \left ( \mathbf{x}  \right ) $ is an NK landscape fully specified  by eq.\ (\ref{NK_1}). 
The additive term is given by
\begin{equation}\label{NK_3}
 F \left ( \mathbf{x}  \right ) = \mathcal{F}_{max}  + \sum_{j=1}^{N} w_j  \left ( \mathbf{x} \right )
 \end{equation}
where $\mathcal{F}_{max} = \mathcal{F}  \left ( \mathbf{x}^{max}  \right ) $ is the fitness value of the unique global maximum $\mathbf{x}^{max}$ of the landscape and 
\begin{equation}\label{NK_4}
w_j \left ( \mathbf{x} \right )=\left \{
			\begin{array}{c l}
                          0  & \mbox{if $x_j = x_j^{max} $}. \\
                     v_j < 0 &  \mbox{if $x_j \neq x_j^{max} $}.
            \end{array}
            		\right.
 \end{equation}
Hence,  once $ \mathcal{F}_{max}$ is known,  we  need only to specify the   $N$ unknowns $v_j$ to determine the values of the additive fitness $F \left ( \mathbf{x}  \right ) $  for all the $2^N$ strings. This can be done  by  considering the $N$ neighbors of the global maximum, $\mathbf{x}_j^{max} = \left (x_1^{max}, \ldots, 1 - x_j^{max}, \ldots, x_N^{max} \right )$ for which we can rewrite
eq. (\ref{NK_3}) as 
\begin{equation}\label{NK_5}
 F \left ( \mathbf{x}_j^{max}  \right ) = \mathcal{F}_{max}  + v_j .
 \end{equation}
Now, since the NK landscape defined in eq.\ (\ref{NK_1}) is clearly additive for $K=0$, we must require that $F \left ( \mathbf{x}  \right )  = \mathcal{F} \left ( \mathbf{x}  \right ) $  in this case. This  condition is fulfilled  provided we define
\begin{equation}\label{NK_6}
 v_j  = \mathcal{F} \left ( \mathbf{x}_j^{max}  \right ) - \mathcal{F}_{max}  
 \end{equation}
for  $j=1,\ldots, N$. So the fitness of the additive landscape  $F \left ( \mathbf{x}  \right ) $ is guaranteed to coincide with the fitness of a general NK landscape $\mathcal{F} \left ( \mathbf{x}  \right )$ at the global maximum and at its neighbors. In addition, in the case $K=0$  we have $F \left ( \mathbf{x}  \right )  = \mathcal{F} \left ( \mathbf{x}  \right ) $ for all strings, since 
$w_j \left ( x_j \right ) = \left [ f_j \left ( x_j \right ) - f_j \left ( x_j^{max} \right ) \right ]/N$.

Let us consider a  trajectory on an NK landscape $\mathcal{F} \left ( \mathbf{x}  \right )$  of length  $T$ that is comprised of the sequences  $\mathbf{x}^{1}, \mathbf{x}^{2},\ldots, \mathbf{x}^{T} =  \mathbf{x}^{max}$. The roughness $\rho$  of the landscape measured along  this trajectory is defined as \cite{Aita_01}
\begin{equation}\label{NK_7}
\rho = \left [  \frac{1}{T}  \sum_{\alpha=1}^T \left  [ \mathcal{F} \left ( \mathbf{x}^\alpha  \right ) - F \left ( \mathbf{x}^\alpha \right ) \right ]^2 \right ]^{1/2} , 
\end{equation}
which depends on the particular trajectory followed by the imitative search on its way towards  the global maximum of the landscape.  The  global  roughness of a landscape $\hat{\rho}$  is defined  by  summing over all the state space in eq. 
(\ref{NK_7}) and setting $T$ to the size of that space, i.e.,  $2^N$.  Notice that we use the term ruggedness in a qualitative sense to mean the jaggedness of the landscape, whereas the term roughness is reserved for the quantity  defined in eq.\ (\ref{NK_7}).

In this paper we focus mainly on single realizations of  NK-fitness landscapes  with $N=12$ for  $K=0,2,4$ and  $11$, so each landscape has $4096$ strings. As expected, the number of maxima increases with $K$, and for the realizations considered we find $1$, $15$, $47$ and  $292$ maxima for $K=0,2,4$ and $11$, respectively. Different realizations yield qualitatively similar results (see, e.g. \cite{ Fontanari_15}).  The  somewhat low dimensionality of the NK landscapes, as well as the use of single landscape realizations,  is justified by  the high computational demand to store and analyze the learning trajectories needed in the calculation of the predictability and the mean path divergence.  We  recall that the number of  trajectories are given by combinations of the number of strings $2^N$ and so, for large $N$, no feasible  ensemble of trajectories can suitably describe the properties of the full trajectory space.  The predictability measure is more susceptible to the increase of the dimensionality of the landscape as it entails the generation of identical trajectories, which  requires an unrealistically large number of  runs. In fact, previous experimental and computational studies have considered landscapes of even lower dimensionality, with the string lengths varying from $N=4$ to $N=8$ \cite{Carneiro_10,Lobkovsky_11,Lobkovsky_12,Visser_14,Aita_01}.  We address the effects of increasing the dimensionality of the landscape  in Section \ref{sec:res}.

\section{Imitative Learning}\label{sec:model}

We consider a well-mixed population of $M$ agents or binary strings of length $N=12$ that explore the state space of an  NK-fitness landscape aiming at reaching the  global maximum
$\mathbf{x}^{max} = \left (x_1^{max},x_2^{max}, \ldots, x_N^{max} \right )$. Initially, all strings are identical (isogenic population) and set to the antipode of $\mathbf{x}^{max}$, i.e., $\left (1-x_1^{max},1 - x_2^{max}, \ldots,1-x_N^{max} \right )$. The initial isogenic population  is necessary because  the predictability measure (see Section \ref{sec:predic})  requires counting the number of identical trajectories in the ensemble of learning trajectories and so all searches must begin and end at the same strings, otherwise they could never be identical. 
The imitative search strategy is based on the presumption that  it may be advantageous for an agent  to copy or imitate the agent with the highest fitness in the population, the so-called model agent \cite{Rendell_10,King_12}. Notice that in this paper  we  use  the terms agent and string interchangeably.

More pointedly, we implement the synchronous or parallel update of the $M$ agents as follows.
At time $t$ we  first determine the model agent and then we let  each agent  choose between two actions. The first action,
which happens with probability $1-p$,   consists of simply flipping a  bit  at random.
The second action, which happens with probability $p$, is the  imitation  of the model agent. The  model and the target agents  are compared  and the different bits are singled out.  Then the target agent   selects at random one of the distinct bits and flips it so that this bit is now the same in both agents.  As expected, imitation  results in  the increase of the similarity between the target and the model agents, which may not necessarily lead to  an increase of the fitness of the target agent if the landscape is not additive.   In the case the target agent is identical to the model agent,  which happens not only in our initial isogenic population setup but  also during the search since the  imitation procedure reduces  the diversity of the strings, the target agent flips  a  bit  at random  with probability one.  After the $M$ agents are updated we identify the  fittest agent in the population, which then becomes the model agent for the next update round at time $t+1$. 

The parameter $p \in \left [0,1 \right ]$ is the imitation propensity, which is the same for all agents (see \cite{Fontanari_16} for the relaxation of this assumption). The case $p=0$ corresponds to the baseline situation in which  the agents explore the state space independently of each other. The case $p=1$ corresponds to the situation where only the model string explores the state space through  random bit flips; the other strings simply follow the model, thus making the population  
almost isogenic. 

We note  that during the increment from $t$ to $t+1$ all    $M$ agents are updated either by flipping a bit at random or by copying a bit of the model string.  The search ends when one of the agents finds the global maximum and we denote by $t^*$ the time when this happens. Clearly, at time $t=t^*$ the model string is the global maximum $\mathbf{x}^{max}$. We stress that  the imitative search always reaches the global maximum since there is a nonzero probability  that a string flips a bit at random,   which guarantees that the agents are always exploring the state space, thus  avoiding  the permanent trapping in the local maxima.  This is true even for $p=1$, since according to the rules of the imitative search, the model string (and only that string)  flips bits at random.    Of course, the time  the imitative search takes to escape the local maxima and reach the global maximum can be very large, as we will see in Section \ref{sec:res}, but it is  definitely  finite.
This  contrasts with the adaptive walks \cite{Kauffman_87}, where only bit flips that increase the fitness of the strings are allowed, which may permanently get stuck in a local maximum of the landscape. For those walks it is important  to study the influence of the topography of the landscape on the  accessibility of the global maximum \cite{Schmiegelt_14}, whereas for the imitative search the relevant issue is determining the time needed to find the global maximum.

In spite of apparent similarities, the imitative learning search is markedly different from the  evolutionary  algorithms (see \cite{Back_96}). In fact, the exploration of the state space  through the flipping of randomly chosen bits  is similar to the  mutation operator of those algorithms, with the caveat that in evolutionary algorithms mutation is an error of the reproduction process, whereas in the imitative search flipping a bit  and imitation are mutually exclusive processes. The analogy between the imitation and the  crossover processes is more flimsy since the model agent is a mandatory parent in all mates but it contributes  a single gene (i.e.,  a single bit) to the offspring which then replaces the other parent, namely, the target agent. Since the contributed gene is not random - it must be absent in the target agent - the genetic  analogy is clearly inappropriate and so
the imitative learning search stands on its own as a search strategy.

The performance of a search is evaluated by the  computational cost $C$  that measures  the  total number of agent updates implemented in a search of duration $t^*$. Since for each round of the search exactly $M$ agents are updated,  we have $ C \propto M t^*$.  In addition, since finding the global maximum of the NK model for $K >0$  is a NP-complete problem \cite{Solow_00}, which means that the time $t^*$ required to solve the
problem  scales with the size $2^N$ of the state space, and in order to guarantee that  $C$ is a quantity on the order of 1 we write
\begin{equation}\label{CC}
C =  M t^*/2^N .
\end{equation}
This definition guarantees that $C$ is on the order of 1 for the independent search (see \cite{Fontanari_15} for an analytical derivation of the computational cost in this case) and, as we will show in Section \ref{sec:res}, for the imitative search as well.
We note that since our searches begin with an isogenic population  set at the antipode of  $\mathbf{x}^{max}$, we expect to obtain  a  higher computational cost as compared with the usual initial setup where the strings are chosen randomly.

Whereas the previous studies of the imitative search focused on the dependence of the computational cost on the two parameters of the search strategy, namely, the number of agents $M$ and the imitation propensity $p$ \cite{Fontanari_14,Fontanari_15}, here our focus is  on  the statistical characterization  of the ensemble of  learning trajectories. We use two quantitative measures  introduced in the study of evolutionary trajectories \cite{Lobkovsky_11} - the predictability and the  path divergence - which we  describe in detail in the next section.

\section{Characterization of  trajectories }\label{sec:predic}

The trajectory of a particular imitative search is the ordered sequence of  the
model strings that begins at the antipode of $\mathbf{x}^{max}$ and ends at $\mathbf{x}^{max}$. Thus an unpurged trajectory has  $t^*$, not necessarily distinct, strings. However, the trajectories we consider here are purged of loops and so they contain  typically much  fewer than $t^*$ strings.
Explicitly, if a string appears as a model string  more than once in a search, thus forming a loop structure, the trajectory is redefined and the loop  is removed. In this way  no string can appear more than once in the resulting purged learning trajectory.   Therefore, for our purposes, a learning trajectory
is described  by a directed graph with no loop structures and where  each node in the graph denotes a model string. We stress that the evaluation of the computational cost (\ref{CC})  does not involve the purging of loops from trajectories, as $t^*$ accounts for  all  (not necessarily distinct) model strings produced in  the search. The loops  need to be purged from the learning trajectories for the calculation of the predictability measure only. In fact,  since in our searches we never found two  identical unpurged trajectories, that measure is not suitable to characterize the ensemble of the  original trajectories, hence the need to purge the loops from them.

For each run of the imitative search,  we keep track of the entire (unpurged) learning trajectory comprising $t^*$ strings  beginning at the antipode of  $\mathbf{x}^{max}$ and ending at $\mathbf{x}^{max}$. We then purge the loops from the trajectory and store it in the trajectory ensemble. This procedure is repeated $\mathcal{A} = 10^6$ times.
Hence, the trajectory ensemble contains $\mathcal{A}$  learning trajectories of the imitative search for a fixed realization of the  NK-fitness landscape and for  fixed parameters $p$ and $M$.  Let us denote by $q_\alpha$  one such a trajectory and by $O \left ( q_\alpha \right )$  its probability of occurrence  in the  ensemble, which is given by the ratio between the number of times  $q_\alpha$ appears in the  ensemble and the ensemble size $\mathcal{A}$.  The predictability of the imitative search is defined as the probability that two independent runs result in the  same (purged) learning trajectory  \cite{Roy_09} (see also \cite{Lobkovsky_11,Lobkovsky_12,Visser_14}), i.e.,
\begin{equation} \label{P2}
P_2 =\sum_{q_\alpha}  O^2 \left ( q_\alpha \right )
\end{equation}
where the sum is over all distinct trajectories in the trajectory ensemble. This  simple measure of the repeatability of the trajectories  varies from $P_2 = 1/\mathcal{A}$ in the case  all trajectories in the ensemble are different to $P_2 = 1$ in the case there is a single trajectory joining the initial and end points of the search  \cite{Visser_14}.
The  inverse of the predictability,   $1/P_{2}$,  can be viewed as 
the effective number of trajectories that contribute to the imitative search. We note that in the spin-glass literature $P_2$ is  akin to the spin-glass order parameter and $1/P_2$  is like a participation ratio \cite{Mezard_86}.

A drawback of the predictability $P_2$ is that very similar, but not identical, trajectories are counted simply as different trajectories and so this quantity offers no clue on the similarity between the learning trajectories of  the imitative search. 
Since all learning trajectories have the same starting and ending points,  we can use the mean path divergence originally introduced in the  study of evolutionary paths \cite{Lobkovsky_11} to assess the trajectories similarity. The key element here is the divergence  $ d \left ( q_\alpha, q_\beta \right ) $ between trajectories $q_\alpha$  and  $q_\beta$, which  may exhibit very distinct lengths, that  is calculated as follows.  For each string in $q_\alpha$ we measure the Hamming distances to all strings in $q_\beta$ and keep the shortest distance only.  Then $ d \left ( q_\alpha, q_\beta \right ) $ is defined as the average of these shortest Hamming distances over all strings in $q_\alpha$. Note that $d \left ( q_\alpha, q_\alpha \right ) = 0$ and $ d \left ( q_\alpha, q_\beta \right )  \neq  d \left (  q_\beta , q_\alpha \right ) $.   Finally,  the mean path divergence is defined as
\begin{equation}\label{d}
  \bar{d}= \sum_{q_\alpha} O \left ( q_\alpha \right ) \sum_{ q_\beta} O \left ( q_\beta \right ) d \left ( q_\alpha, q_\beta \right ),
\end{equation}
which yields the expected divergence of two trajectories  drawn at random from the ensemble of trajectories  \cite{Lobkovsky_11}. We note that the time (or the position of the strings in the purged trajectory) factors out of the divergence  $ d \left ( q_\alpha, q_\beta \right )$. This is so because the trajectories typically have very distinct lengths,  which limits the usefulness of   comparing  strings  at the same position in both trajectories. In addition,  since we purged the loops from the trajectories the time information is lost, although the order of the strings is maintained.

\section{Results}\label{sec:res}

As a measure of  the performance of the population in searching for the global maximum of the NK-fitness landscapes, we consider the mean  computational cost $\langle C \rangle $, which  is obtained by averaging the computational cost (\ref{CC}) over $\mathcal{A} =10^6$  searches for each  landscape realization.  The dependence of this mean cost on the model parameters $M$ and $p$ for  homogeneous, well-mixed   populations has been extensively studied in previous works \cite{Fontanari_15}. 

\begin{figure}  
\centering  
 \includegraphics[width=0.48\textwidth]{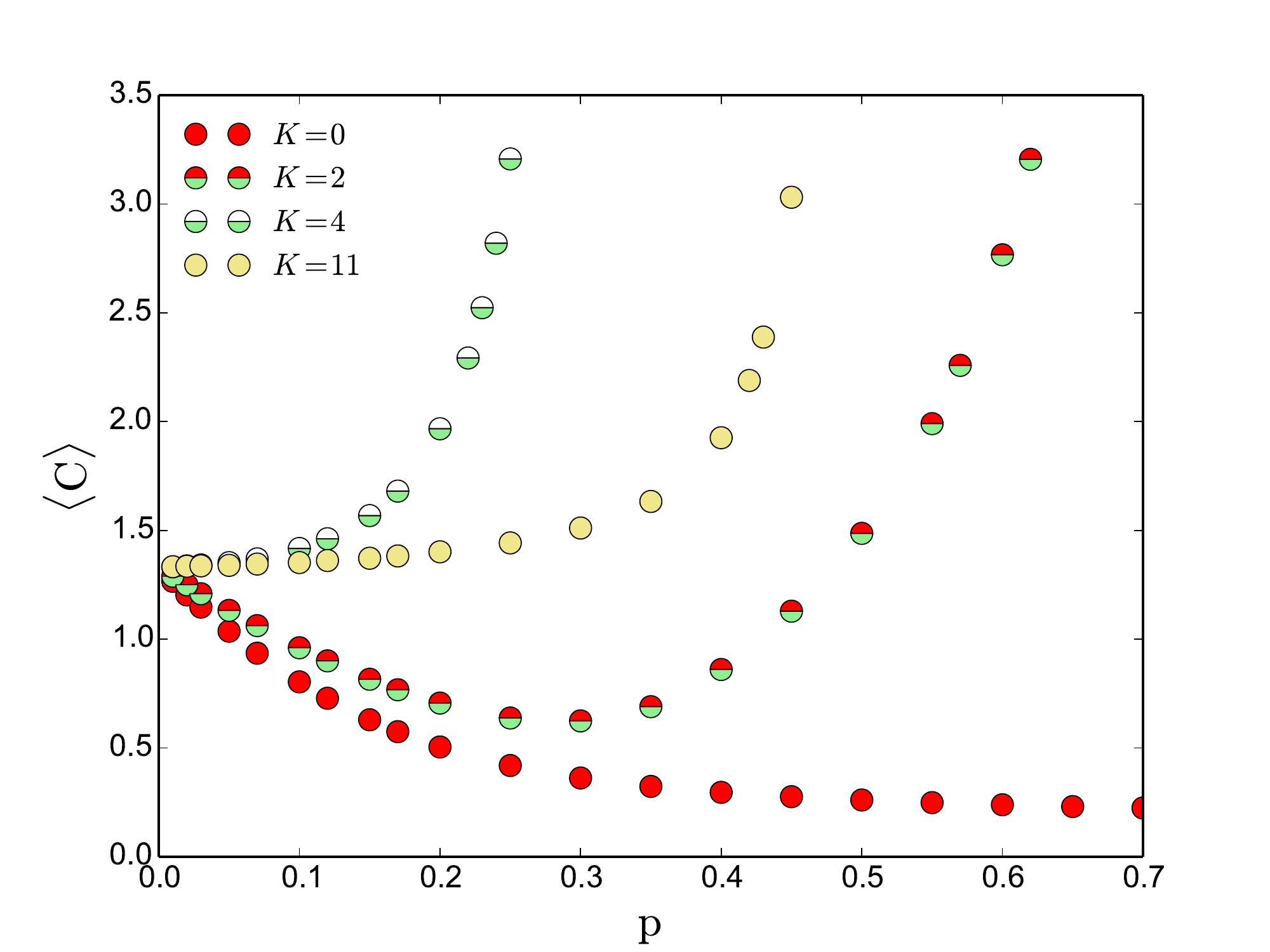}  
\caption{Mean computational cost $\langle C \rangle $ as function of  the imitation propensity $p$ for populations of fixed size $M=50$. The parameters of the fitness landscape are  $N=12$ and  $K=0, 2, 4, 11$ as indicated.}  
\label{fig:fig1}  
\end{figure}

Figure \ref{fig:fig1} shows $\langle C \rangle $ against the imitation propensity $p$ for populations of fixed size $M=50$ and for landscapes of distinct ruggedness. For the smooth, additive landscape ($K=0$) the cost decreases monotonously with increasing $p$. This is expected because in this case the fitness of a string is highly correlated to its distance to the global maximum and so it is always beneficial to imitate the  model agent. This scenario changes for rugged landscapes due to the presence of local maxima that may temporarily trap the entire population if the imitation propensity is too high.  The  catastrophic performance  observed in this case   is akin to the Groupthink phenomenon \cite{Janis_82},  when everyone in a group starts thinking alike,  which can occur when people put unlimited faith in a talented leader (the model agent). 
 In this case, there is an optimal value of the imitation propensity that minimizes the cost. For too rugged landscapes (say, $K \geq 4$ in Fig.\ \ref{fig:fig1}), the best strategy  is never imitate the model (i.e., $p=0$),   thus allowing   the agents  to explore the landscape independently of each other. 
 
 %-----------------------------------------------------
%-----------------------------------------------------
  
  Figure \ref{fig:fig1}  reveals a curious result: the hardest challenge to the imitative search is not the most rugged landscape ($K=11$)  but   the moderately rugged landscape ($K=4$).  In fact, the common notion that the hardest challenges to search heuristics  are landscapes with a large number of local maxima (i.e., $K=N-1$) stems from the study of adaptive walks, which are prone to get permanently stuck in those  maxima \cite{Kauffman_87,Kauffman_95}. Different search heuristics, however, may be affected by other features of the landscape. For instance, the imitative search is more affected by the local maxima for $K=4$ being  farther apart than for $K=11$, which makes the escape from them much more costly. 
  
  %-----------------------------------------------------
\begin{figure}  
\centering  
 \includegraphics[width=0.48\textwidth]{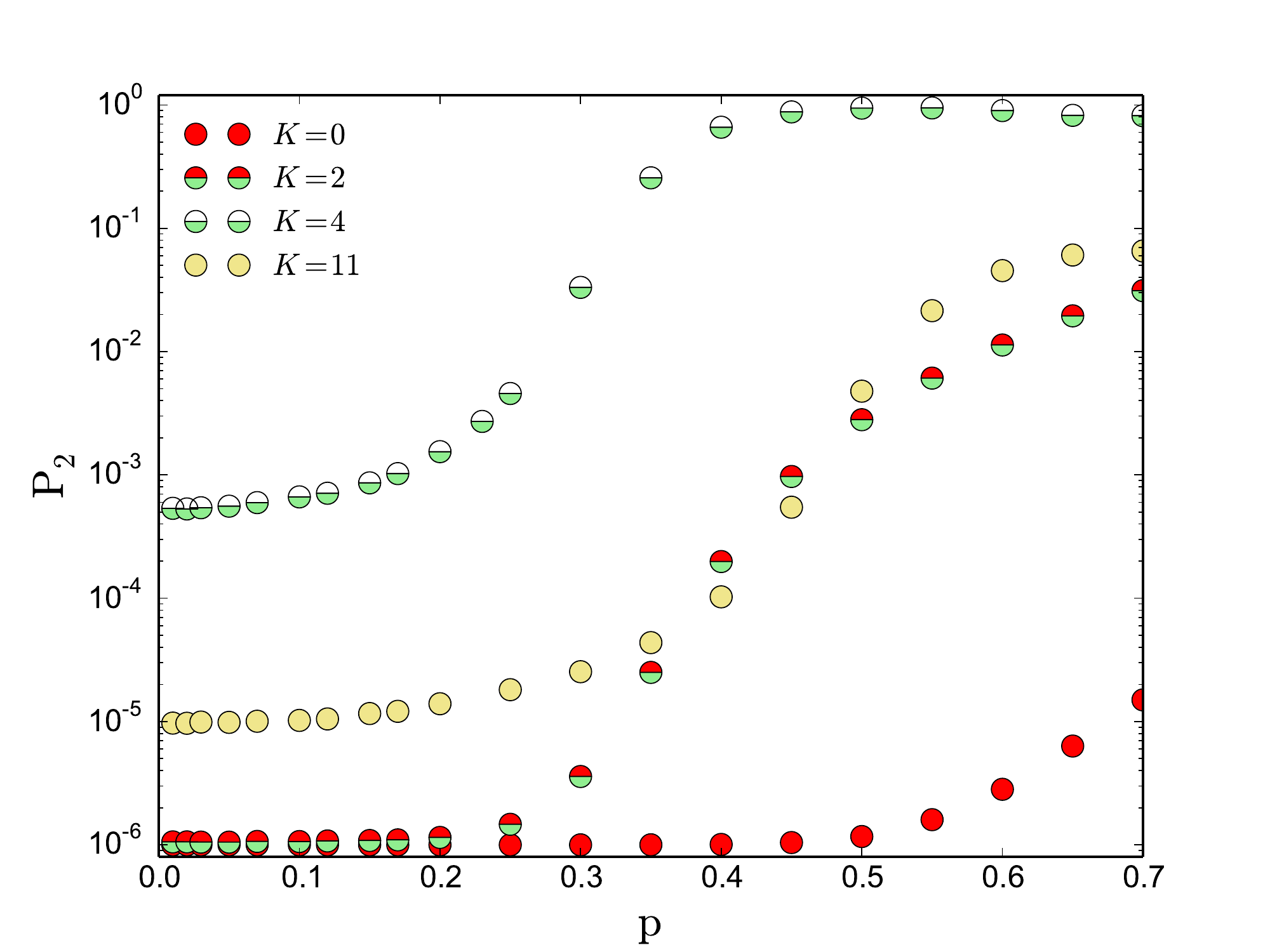}  
\caption{Predictability $P_2$ of the learning trajectories  as function of  the imitation propensity $p$ for populations of fixed size $M=50$. The parameters of the fitness landscape are  $N=12$ and  $K=0, 2, 4, 11$ as indicated.}  
\label{fig:fig2}  
\end{figure}
%-----------------------------------------------------

We  turn now to the study of the learning trajectories using the two measures introduced in Section \ref{sec:predic}.  Figure \ref{fig:fig2} shows the predictability $P_2$ for the learning trajectories that resulted in the mean computational costs exhibited in 
Fig.\ \ref{fig:fig1}.  The results of Fig.\ \ref{fig:fig2} are bewildering since they show that even if the computational costs are practically indistinguishable for the different landscapes, as in the case of small $p$, the ensembles of learning trajectories are completely different.   In particular, for the smooth ($K=0$) or the nearly smooth ($K=2$)  landscapes, for which the local maxima have little effect on the search (see Fig.\ \ref{fig:fig1}), the learning trajectories are practically unpredictable: each run follows a distinct learning trajectory.  For the more rugged landscapes, however, the local maxima seem to act as mandatory rest areas of the imitative search resulting in a substantial increase of the predictability. This effect is enhanced by the increase of the imitation propensity, as expected, making the dynamics almost deterministic for the landscape with $K=4$. 

%In contrast to the computational cost (Fig.\ \ref{fig:fig1}),   the ensemble of trajectories shows a strong dependence  on the ruggedness of the landscape even in the case $p=0$ (independent search). This is so because the learning trajectories are the trajectories of the model strings, i.e., the strings of highest fitness at a given time, and so the evolution of those strings must necessarily be influenced by the  fitness landscape. In this case, the attractivity of the local maxima is due to the effect of reverse bit flippings. 

\begin{figure}  
\centering  
 \includegraphics[width=0.48\textwidth]{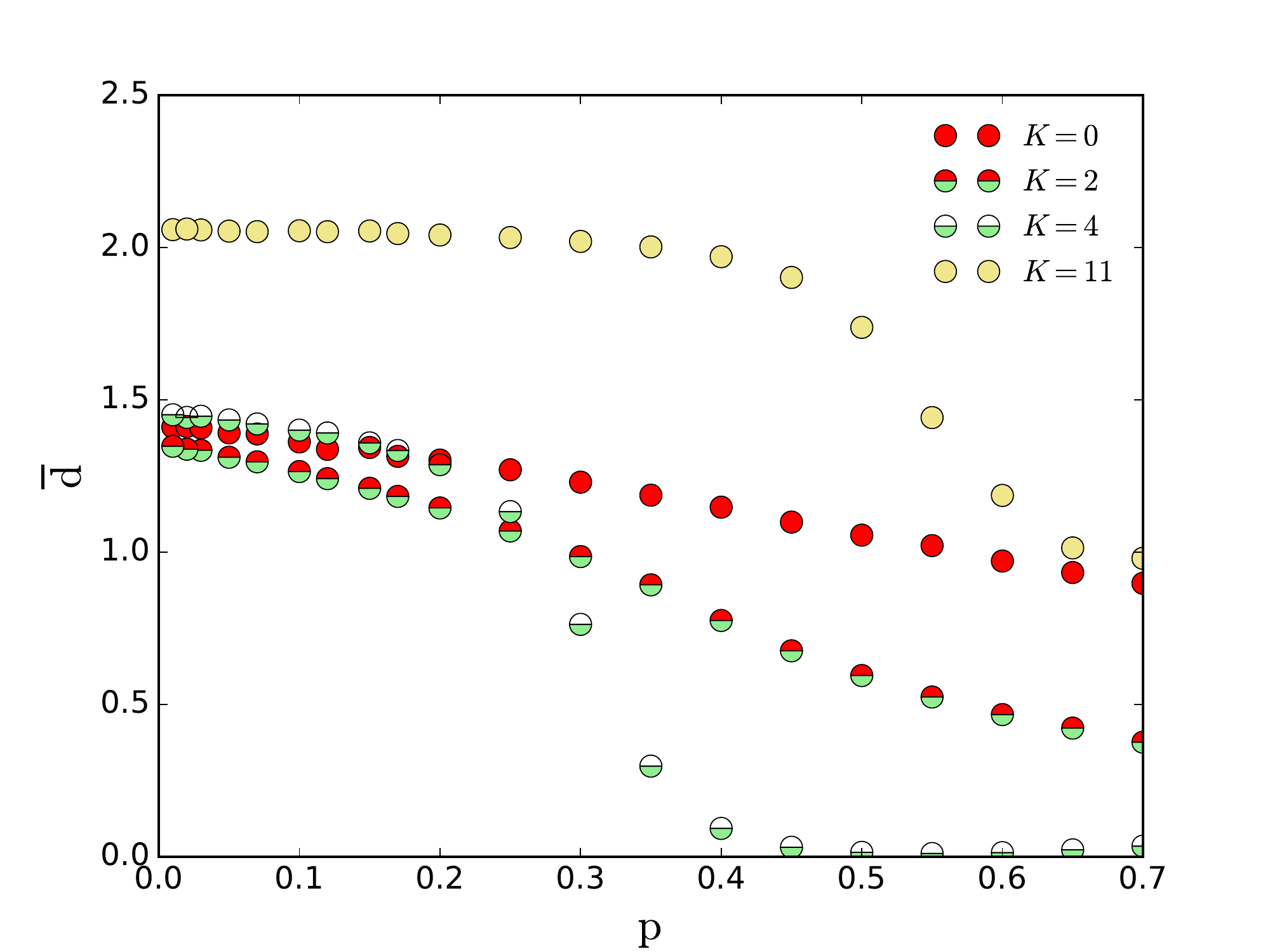}  
\caption{Mean path divergence $\bar{d}$ of the learning trajectories as function of  the imitation propensity $p$ for populations of fixed size $M=50$. The parameters of the fitness landscape are  $N=12$ and  $K=0, 2, 4, 11$ as indicated.}  
\label{fig:fig3}  
\end{figure}

As pointed out in Section \ref{sec:predic}, the mean path divergence $\bar{d}$ offers a view of the similarity of  the  learning trajectories. This quantity is shown in Fig.\ \ref{fig:fig3} for the same trajectories considered in the analysis of the predictability. The monotonic decreasing of $\bar{d}$ with increasing $p$ is expected since the overlap among trajectories must increase due to  the boosted influence of  the local maxima.  We note that there is no direct relation between the predictability $P_2$ and the path divergence $\bar{d}$. For instance, consider a situation where the search can follow   two  trajectories only  and that those, say equally probable, trajectories have  negligible overlap. This  hypothetical scenario would produce  highly divergent and highly  predictable trajectories.   Thus, although there are fewer distinct trajectories in the ensemble for $K=11$ than in the ensemble for $K=0$ (see Fig.\ \ref{fig:fig2}), the  distinct trajectories for $K=11$ are much more divergent than those for $K=0$.  In this line, the comparison between the results for $K=4$ and $K=0$ for small $p$ indicates that the few distinct trajectories that  compose the ensemble of trajectories  for $K=4$   are as divergent as the multitude of trajectories of the ensemble for $K=0$. However, for high $p$ the ensemble of trajectories  for $K=4$ exhibits the hallmarks of a deterministic dynamics, namely, high predictability and low divergence.

%-----------------------------------------------------
%-----------------------------------------------------

A word is in order about the effect of the size of the ensemble of trajectories  (i.e., the number of searches) $\mathcal{A} = 10^6$  on our results. This ensemble size is more than enough for a high accuracy estimate of the mean computational cost (Fig.\ \ref{fig:fig1}) and the mean path divergence (Fig.\ \ref{fig:fig3}). In fact, a much smaller number of searches (e.g., $10^4$) yields the  same results for those quantities.  However, a very  large $\mathcal{A}$  is required to  accurately  estimate  the predictability $P_2$. We recall  that this quantity can be interpreted as the probability that two identical  trajectories  are randomly selected  with replacement from our ensemble of trajectories (see eq.\ (\ref{P2})). Since  the minimum value  that $P_2$ can assume  is $1/\mathcal{A}$, our estimate is reliable provided that 
$P_2  \gg 1/\mathcal{A}$, which guarantees a significant number of identical trajectories in the ensemble.    Hence for the data shown in Fig.\ \ref{fig:fig2}, our results are not  reliable quantitatively in the regions where $P_2$ is on the order of   $1/\mathcal{A}= 10^{-6}$. Qualitatively, however, they correctly point out the very low predictability of the search in those regions.

The previous analysis considered a fixed population size $M=50$ and focused on the influence of the imitation propensity $p<0.7$  on the
properties of the ensemble of learning trajectories.  We recall that the number of agent updates the imitative search requires to find the global maximum is given by $2^{N} C$  (see eq.\ (\ref{CC})) and so for $N=12$ the search becomes prohibitively demanding   if $C > 100$.  Since for $K > 0$ the cost increases steeply with increasing $p$ (see Fig.\ \ref{fig:fig1}), it is computationally unfeasible  to carry out searches beyond $p=0.7$.  For very short strings, say $N=5$, we can easily run searches with $p=1$ because a high computational cost, say $C$ on the order of $10^4$, does not correspond to a too large number of updates.  Actually, this is how   we know that the computational cost does not diverge for $p=1$. 
Of course, this is not an issue for the additive landscape ($K=0$), since in this case the  computational cost decreases monotonically with increasing $p$.

\begin{figure}  
\centering  
 \includegraphics[width=0.48\textwidth]{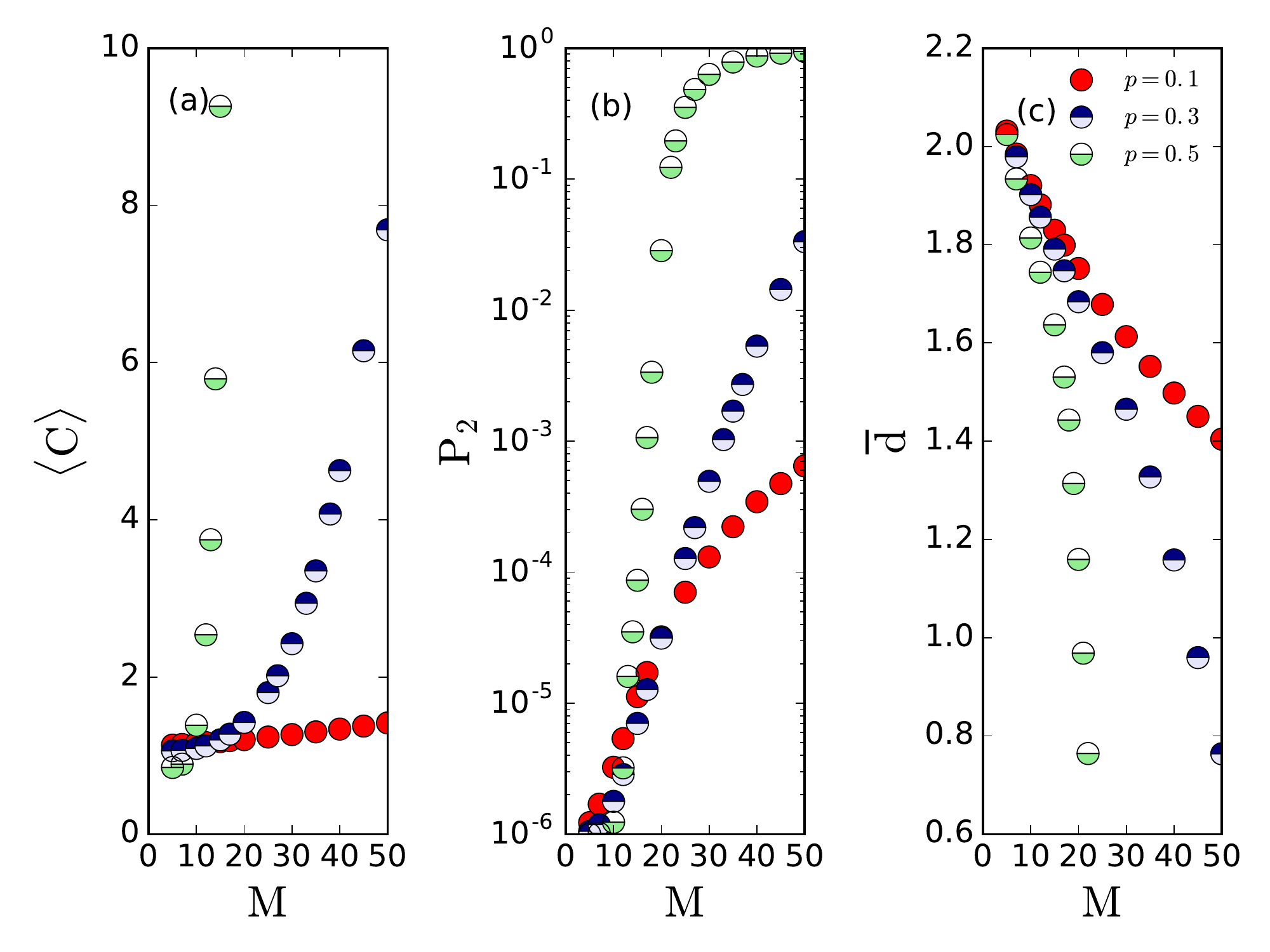}  
 \caption{Influence of the population size $M$ on  (a) the mean computational cost $\langle C \rangle $, (b)  the predictability $P_2$  and (c)  the mean path divergence  $\bar{d}$,  for the imitation propensities $p=0.1, 0.3, 0.5$ as indicated. The parameters of the landscape  are $N=12$ and $K=4$.}  
\label{fig:fig4}  
\end{figure}

   Regarding  the  predictability of stochastic dynamic  processes, the study of the  influence of the population size $M$ is important as one usually expects that the dynamics becomes more deterministic as the population increases.
Figure \ref{fig:fig4}  summarizes our findings for a fixed NK-fitness landscape with $N=12$ and $K=4$. Both measures $P_2$ and $\bar{d}$ indicate that the learning trajectories become more deterministic, in the sense that there are fewer distinct  trajectories and they are more similar to each other, as $M$ increases. The reason for that is the strengthening of the attractivity of the local maxima as evinced by the increase of the computational cost. In fact, if the  search is forced to visit the regions around those maxima, the resulting  trajectories must exhibit a high overlap. Large populations increase the attractivity of the local maxima simply because they allow the existence of several copies of the model agent and
  this makes it  very hard for the imitative search to explore other regions of the state space through the flipping of bits at random, since the extra copies attract the updated model agent back to the local maximum. Hence what makes the dynamics more deterministic with increasing $M$  is not the decrease of the size of fluctuations around the deterministic trajectory, but the mandatory passage through the local maxima of the landscape and the few routes of escape from them.

%-----------------------------------------------------
%-----------------------------------------------------

%-----------------------------------------------------
\begin{figure}  
\centering  
 \includegraphics[width=0.48\textwidth]{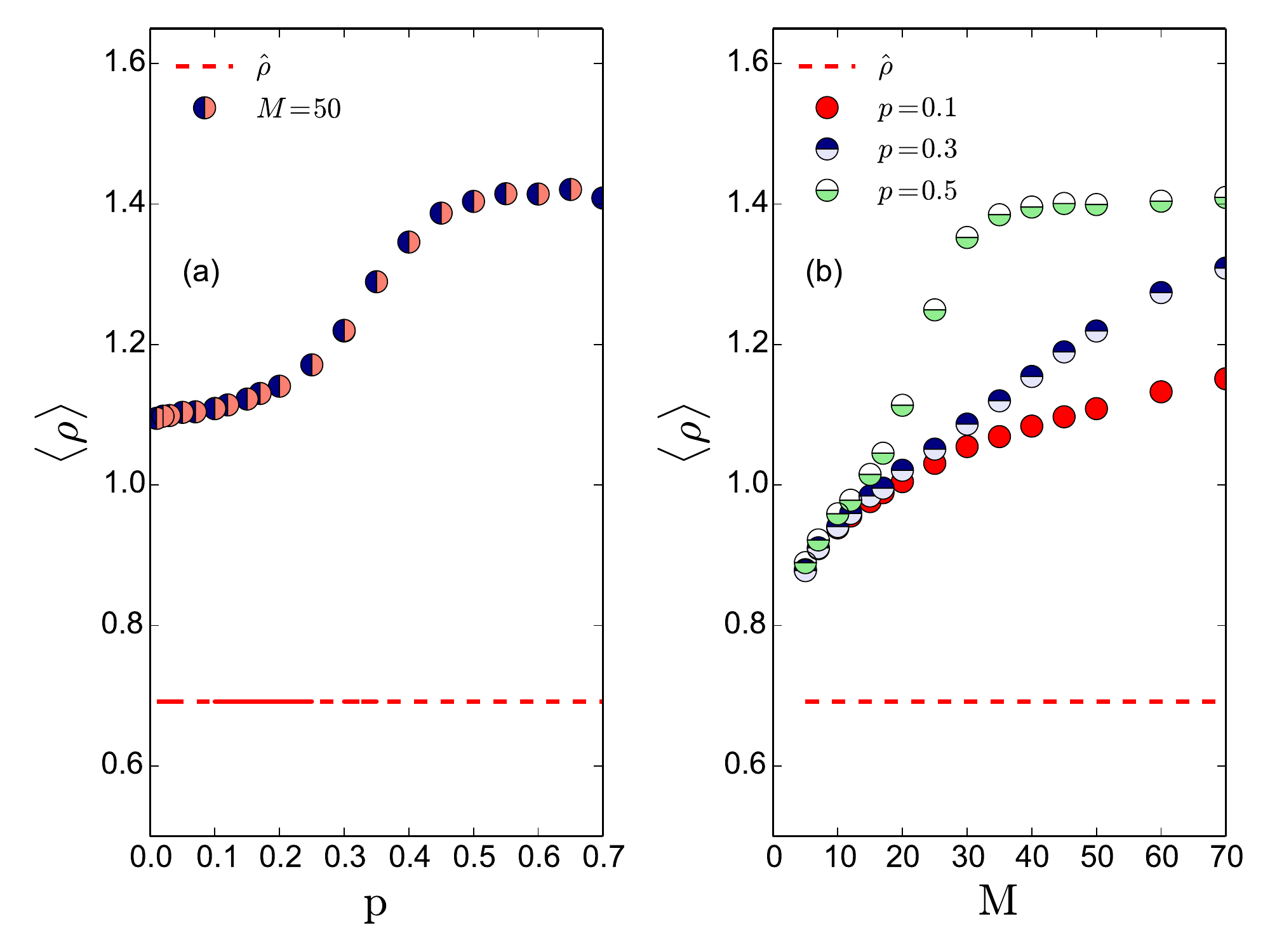}  
 \caption{Mean roughness $\langle \rho \rangle $ of the learning trajectories   as function of (a) the imitation propensity $p$ for $M=50$ and (b) the population size $M$ for $p = 0.1, 0.3,  0.5$  as indicated.  The dashed lines show the global roughness of  the landscape, $\hat{\rho} = 0.6917$. The parameters of the landscape  are $N=12$ and $K=4$.}  
\label{fig:fig5}  
\end{figure}
%-----------------------------------------------------

Finally, Fig.\ \ref{fig:fig5}  reveals the  utility of the trajectory-dependent roughness $\rho$, which is defined in eq.\ (\ref{NK_7}), in elucidating  the nature of the imitative learning search. Actually, since $\rho$ is defined for a particular trajectory, this figure presents the average of $\rho$ over the $\mathcal{A}=10^6$  trajectories of the ensemble of learning  trajectories. The results show that the imitative search  seems to experience a more rugged landscape as $p$ or $M$ increases, although the landscape is  fixed for the data shown in the figure.  
We recall that the roughness  $\rho$ of a given trajectory on a rugged landscape measures the deviation from additivity of the fitness values of  the strings that compose the trajectory  \cite{Aita_01}.  Since the actual realizations of the learning trajectories are determined by the rules of the imitative search,  Fig.\  \ref{fig:fig5} shows that  the search follows trajectories that depart markedly from additivity in the cases of large $M$ and $p$, i.e., the imitative search seems to bias the agents towards the   roughest paths that connect the global maximum to its antipode.   For very large $M$ or $p$ close to 1, the roughness  seems to saturate to a value that depends only on the landscape and that characterizes the trajectory of maximum  roughness  on  it.
 Moreover, regardless of the values of the  parameters  $M$ and $p$, the roughness of the learning  trajectories  are always greater than the  global roughness   $\hat{\rho}$ of the landscape, as indicated in Fig.\  \ref{fig:fig5}. The reason is that the imitative search is attracted to regions around the local maxima, where the rugged landscape differs the most from its additive counterpart.  Since the computational cost is very high in the regime where the roughness is large (see Fig.\ \ref{fig:fig4}), our results hint that an efficient search strategy should follow the smoothest paths to the global maximum. This leads to the interesting question of determining the minimum possible roughness of a landscape. In fact, in the same sense that the global roughness  $\hat{\rho}$ can be seen as a lower bound for the maximum roughness path, it is an upper bound for the minimum roughness path. The determination and characterization of those two extreme paths may offer a promising alternative approach to  study rugged fitness landscapes.

 To conclude, we address now  the effect of increasing the dimensionality of the landscape, which is determined by the  length $N$ of the strings.  This can be understood  by considering the correlation between the fitness of any two neighboring strings, which is given by $ corr \left ( \mathbf{x}, \tilde{\mathbf{x}}_i \right ) = 1 - \left ( K+1 \right )/N $,  where $\tilde{\mathbf{x}}_i$ is the string  $\mathbf{x} $ with bit $i$ flipped, and the fact  that the mean density of local maxima is given by $1/\left ( N +1 \right)$  for the uncorrelated NK landscape, i.e., for  $K=N-1$  \cite{Kauffman_87}. Another important piece of information to understand the effect of $N$  is the so-called complexity catastrophe, which asserts that, regardless of the value of $K>0$ , the average fitness of the local maxima  equals the average  fitness of the landscape  for large $N$  \cite{Kauffman_87,Solow_99}. Hence for fixed $K$, increase of the dimensionality of the landscape makes it more correlated and  the local maxima sparser.
In addition, since the fitness of those maxima are only marginally greater than the mean fitness of the landscape for large $N$,  the imitative search can easily escape them. In sum, since high dimensional NK-fitness landscapes are essentially smooth landscapes for which the trap effect of the  local maxima is greatly attenuated, we expect  the learning trajectories  to be highly unpredictable, similarly to  the case $K=0$.

\begin{figure}
\centering  
 \includegraphics[width=0.48\textwidth]{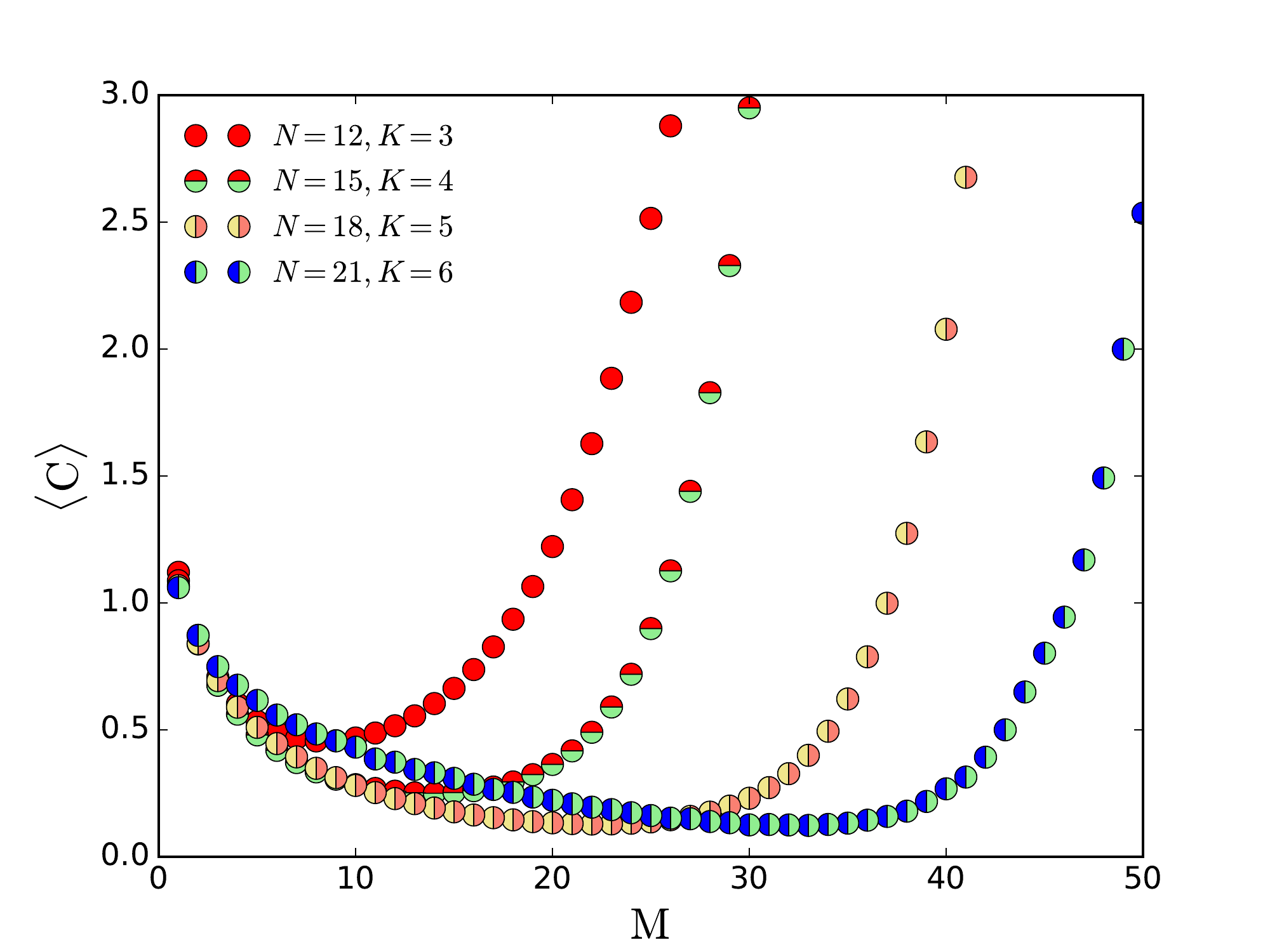}  
\caption{Mean computational cost $\langle C \rangle $  as function of the population size $M$    for  four  families of  NK landscapes $(N=12,K=3)$, $(N=15,K=4)$, $(N=18,K=5)$ and  $(N=21,K=6)$ as indicated.   The mean number  of local maxima is  $28.7$, $125.0$, $605.2$ and $3115.5$, respectively, and the  imitation propensity is $p=0.5$. 
 }
\label{fig:fig6}
\end{figure}

The smooth landscape scenario for high dimensional NK landscapes is corroborated by  an analysis of the computational cost of NK landscapes with increasing $N$
and fixed local fitness correlation. Figure \ref{fig:fig6} shows this study for landscapes with   $(N=12,K=3)$, $(N=15,K=4)$, $(N=18,K=5)$ and $(N=21,K=6)$, which exhibit the same local correlation, viz. $ corr \left ( \mathbf{x}, \tilde{\mathbf{x}}_i \right ) =2/3$.  As this analysis focus on the computational cost only, we run $10^4$ searches for each landscape. In addition, for each set of the parameters $N$ and $K$ we generate and store $100$ landscape realizations, so as to guarantee that runs  with different search parameters ($M$ in the case of the figure) explore exactly the  same landscape realizations. The results of Fig.\  \ref{fig:fig6}, where each point represents an average over $10^4$ searches and $100$  landscape realizations,   show that the computational cost tends to  a monotonically decreasing function of  $M$  when the length of the strings $N$ becomes very large, which is what one expects of a search on a landscape  unplagued by local maxima.
 %
 %----------------------------------------------------

\section{Discussion}\label{sec:disc} 
The  study of the statistical properties of  the ensembles of  trajectories  offers 
 an alternative  perspective to assess the behavior of agent-based models, whose  analysis has mostly  focused  on the  length and on the nature of the  ending points of the dynamic trajectories.  Although the original motivation to study  ensembles of evolutionary trajectories was to  offer an answer to the ultimate (and still open) question  whether evolution is deterministic and hence predictable or   stochastic and hence unpredictable \cite{Lobkovsky_11,Visser_14}, the measures for the  predictability and the divergence of trajectories  introduced in those studies could also be fruitfully applied to study dynamical systems outside the biological realm.

Accordingly, in this paper we have analyzed the learning trajectories of the imitative search \cite{Fontanari_14} on NK-fitness landscapes using the predictability measure $P_2$, which yields the probability that two independent runs of the search  follow the same trajectory on the state space \cite{Roy_09}, and the path divergence $\bar{d}$, which measures the dissimilarity of the  learning trajectories \cite{Lobkovsky_11}.  Regarding the use of these measures,  a  great advantage of the imitative search  over the traditional algorithms used to model the evolutionary dynamics is the ease to define a learning trajectory, which is the ordered, loop-free sequence of model strings that
ends at the (unique) global maximum  of the landscape and begins at its antipode. Since in the imitative search the model strings play 
an important role in guiding the population towards the global maximum, it is natural to use them to compose the learning trajectories.
In the evolutionary algorithms, however, there is no such  natural choice: the fittest string at a given generation is more likely to contribute offsprings to the the next generation but does not have a global guiding role as the model string of the imitative learning, and the consensus string is an artificial construct that plays no role in the population dynamics.

We find that the measures $P_2$ and $\bar{d}$ depend not only on the underlying  fitness landscape but also on the way the population explores the landscape, which  may enhance some features of its topography as, for instance, the   attractivity of the local maxima. This effect makes it difficult to draw general conclusions on the influence of the topography on the learning trajectories. Nevertheless, our results for  the predictability of the trajectories (see Fig.\ \ref{fig:fig2}) accord with the empirical findings that the landscape ruggedness   caused by magnitude and sign epistasis  constrains  the pathways of protein evolution without shutting off pathways to the maximum  \cite{Carneiro_10}. This is exactly what  we observe when we compare the predictability for the additive ($K=0$) landscape with the predictability for rugged landscapes. This result is expected since a rugged landscape exhibits a multitude of minima that are actively avoided by both the imitative and the evolutionary  searches. Avoiding these minima as well as  the regions of low fitness around them, restricts the learning trajectories and hence increases the predictability. However, the predictability exhibits a  non-monotonic dependence on the ruggedness  of the NK-fitness landscapes: although the  trajectories on a rugged landscape are more predictable than  on a smooth landscape, increasing the ruggedness of the landscape does not necessarily increase the  predictability of  the trajectories.

The  high   predictability and low  divergence trajectories observed when  the  population size $M$ or the  imitation propensity $p$ are large  correlate very well with the regions of  high computational cost, which  indicates that these trajectories were trapped in local maxima, the escape of which was very costly to the imitative search. We note that the reason the local maxima makes the learning trajectories more deterministic is not only that they attract the trajectories but that there are only a few effective escape routes away from them. This perspective allows us to understand why the computational cost and the predictability are higher for the landscape with $K=4$ than for the uncorrelated landscape with $K=11$ (see Figs.\ \ref{fig:fig1} and \ref{fig:fig2}):  although the latter landscape exhibits more local maxima, there are many effective and distinct escape  paths from them. The mean path divergence (see Fig.\ \ref{fig:fig3}) accords with  and adds to this scenario by showing that the escape paths are more divergent for the uncorrelated landscape. In sum, we conclude that high levels of determinism come at the cost of computational effort. We note, however, that there is no gain on the learning trajectories being more (or less) deterministic, and that the imitative search is inherently a stochastic search heuristic, despite the high predictability of the learning trajectories for some parameters of the search.

In addition to the analysis of the predictability and divergence measures,  the knowledge of the learning trajectories allows us to examine the influence of the model parameters on the way the imitative search experiences the landscape. This is   possible due to the introduction  of the trajectory-dependent roughness $\rho$, defined in eq.\ (\ref{NK_7}), that reveals quite explicitly that the imitative search follows qualitatively distinct trajectories as the control parameters  $M$ and $p$ change. As pointed out, large values of these parameters increase the attractivity of the local maxima, which results in an increase of  the roughness of  the learning trajectories.  In fact, since $\rho$ measures  essentially the deviation from additivity \cite{Aita_01} and the local maxima are the hallmarks of a non-additive landscape, a trajectory  that visits a large number of them (or their vicinities) is guaranteed to maximize the roughness.

Although our study focused solely on time-independent fitness landscapes, we mention that  in the contexts of   cultural evolution and social learning time-dependent fitness landscapes  are of utmost relevance (see, e.g., \cite{Wilke_99} for the study of adaptive walks on  time-dependent NK landscapes),  since it is precisely the variability of the environment that promotes the evolution of learning, or more generally,  of phenotypic plasticity \cite{Scheiner_17}. We note, however, that the whole issue of the predictability of evolution presupposes that the environment  is either constant, as considered here, or that it changes in the same way  each time the evolution tape is played (i.e., the landscape changes  deterministically),  otherwise the notion of predictability would make little sense \cite{Gould_97,Morris_10}.

In view of the widespread use of the fitness (or adaptive) landscape metaphor on so many distinct branches of science \cite{Kauffman_95},  the analysis of the  ensemble of trajectories using the tools  discussed in this paper is likely to  greatly  impact  our understanding of a vast class of dynamical systems. Here we have shown how this analysis sheds light on the  intricate interaction between the  imitative learning search  and the underlying fitness landscape, revealing the crucial role  the local maxima  play as guideposts  on the route towards the global maximum.

\acknowledgments
P.R.A.C. is supported in part by  Conselho Nacional de Desenvolvimento Cient\'{\i}fico e Tecnol\'ogico (CNPq) by Grant No.\ 303497/2014-9 
and Funda\c{c}\~ao de Amparo \`a Ci\^encia e Tecnologia do Estado de Pernambuco (FACEPE) under Project No.\ APQ-0464-1.05/15. J.F.F. is  supported in part by Grant No.\  2017/23288-0, Fun\-da\-\c{c}\~ao de Amparo \`a Pesquisa do Estado de S\~ao Paulo 
(FAPESP) and  by Grant No.\ 305058/2017-7, Conselho Nacional de Desenvolvimento 
Cient\'{\i}\-fi\-co e Tecnol\'ogico (CNPq).

%\bibliographystyle{chicago} 
%\bibliography{Predictability}

\end{document}